\documentclass[aps,pra,twocolumn,superscriptaddress]{revtex4}%
\usepackage{graphics}
\usepackage[mathscr]{euscript}
\usepackage{amsmath}
\usepackage{amsfonts}
\usepackage{amssymb}
\usepackage{epsfig}
\usepackage{hyperref}
\usepackage{braket}
\usepackage{siunitx}
\usepackage{blkarray}
\usepackage{xcolor}

\DeclareMathOperator{\diag}{diag}
\DeclareMathOperator{\tr}{Tr}
\newcommand{\absA}{\mathbb{A}}
\newcommand{\absB}{\mathbb{B}}
\newcommand{\relayQ}{\gamma}
\newcommand{\evePrime}[0]{{\mathfrak{E}'}}
\DeclareMathOperator{\eveStates}{\mathfrak{E}}
\newcommand{\eveStatesPrime}[0]{\mathfrak{E}'}

\newcommand*{\tran}{^{\mkern-1.5mu\mathsf{T}}}
\DeclareMathOperator{\bobGuess}{\tilde{\kappa}}
\newcommand{\relayVarOQ}{\upsilon}
\newcommand{\relayVarOQRE}{\tilde{\upsilon}}

\DeclareMathOperator{\HBRE}{\chi^{\text{RE}}}

\begin{document}
\title{Long-distance continuous-variable measurement-device-independent quantum key distribution with post-selection}
\author{Kieran N. Wilkinson}
\affiliation{Department of Computer Science, University of York,
York YO10 5GH, United Kingdom}
\author{Panagiotis Papanastasiou}
\affiliation{Department of Computer Science, University of York,
York YO10 5GH, United Kingdom}
\author{Carlo Ottaviani}
\affiliation{Department of Computer Science, University of York,
York YO10 5GH, United Kingdom}
\author{Tobias Gehring}
\affiliation{Department of Physics, Technical University of
Denmark, Fysikvej, Kongens Lyngby 2800, Denmark}
\author{Stefano Pirandola}
\affiliation{Department of Computer Science, University of York,
York YO10 5GH, United Kingdom}

\begin{abstract}
We introduce a robust scheme for long-distance continuous-variable (CV) measurement-device-independent (MDI) quantum key distribution (QKD) in which we employ post-selection between distant parties communicating through the medium of an untrusted relay. We perform a security analysis that allows for general transmissivity and thermal noise variance of each link, in which we assume an eavesdropper performs a collective attack and controls the excess thermal noise in the channels. The introduction of post-selection enables the parties to sustain a secret key rate over distances exceeding those of existing CV MDI protocols. In the worst-case scenario in which the relay is positioned equidistant between them, we find that the parties may communicate securely over a range of \SI{14}{\km} in standard optical fiber. Our protocol helps to overcome the rate-distance limitations of previously proposed CV MDI protocols while maintaining many of their advantages.
\end{abstract}

\maketitle

\section{Introduction}
With the promise of provably secure communication built on the laws of physics, Quantum key distribution (QKD) \cite{pirandola2019advances, gisin2002quantum} is one of the most important results emerging from the field of quantum information theory \cite{nielsen2002quantum, weedbrook2012gaussian}. QKD allows two parties, conventionally named Alice and Bob, to generate a secret key by communicating via an untrusted quantum channel. An eavesdropper (Eve) may employ the most robust attack allowed by the laws of physics, however, she is always restricted by the inherent uncertainty of quantum mechanics and is forced to avoid overtampering with the signal as doing so will reveal her presence to the parties. By combining the attained secret key from a QKD protocol with the one-time pad algorithm, fully secure communication between the parties is guaranteed.

In recent years the field of QKD has evolved rapidly from the primitive BB84 protocol \cite{bennett2014quantum} to current state-of-the-art provably secure protocols allowing parties to communicate over hundreds of kilometers \cite{korzh2015provably,gottesman2004security,yin2017satellite}. Furthermore, there exists a large body of work based on proof-of-principle experiments and in-field tests, including ground-to-satellite communications \cite{hiskett2006long, dixon2008gigahertz, liao2018satellite}. Most of the aforementioned work has focused on discrete variable (DV) protocols. Continuous variable (CV) protocols are promising alternatives that make use of readily available, inexpensive, and easily implementable equipment. CV protocols have been demonstrated to be capable of secret key rates close to the ultimate repeaterless (PLOB) bound~\cite{pirandola2017fundamental}, which corresponds to the secret key capacity of the lossy channel. Many protocols have been proven secure and others have been demonstrated in a proof-of-concept experiment
\cite{jouguet2013experimental} and field tests \cite{huang2016field, CVQKDlong1}. Recently, experimental results for long-distance CV QKD over \SI{202.81}{\km} of ultralow-loss optical fiber have been achieved~\cite{CVQKDlong2}.

Many recent QKD protocols have focused on an end-to-end as opposed to point-to-point approach in which Alice and Bob communicate via remote relays. Introducing a single relay allows the parties to perform measurement-device-independent (MDI) QKD protocols, even if the relay is untrusted \cite{braunstein2012side, lo2012measurement, ma2012alternative, wang2013three, zhou2016making}. Measurement device independence removes the security threat of side-channel attacks attempted by Eve. Several MDI-inspired protocols have been devised that can achieve high rates and exceed the PLOB bound. The first of these protocols was the seminal twin-field protocol~\cite{TF1,TF2,TF3}, followed by the phase-matching protocol~\cite{PM1,PM2} and the sending-or-not-sending protocol~\cite{SNS1,SNS2,SNS3,SNS4}. See Fig.~11 of Ref.~\cite{pirandola2019advances} for a summary of their performances.

CV MDI was recently proposed and demonstrated in a proof-of-concept experiment to achieve very high secret key rates over relatively short distances~\cite{pirandola2015high} (see also Refs.~\cite{ottaviani2015continuous, Leo1, Leo2} for other studies). Unfortunately, developing a protocol that allows exploitation of the practicality of the CV MDI regime at long distance is a difficult problem in recent QKD theory~\cite{hard1,hard2,hard3,hard4}. A lot of effort has been directed at improving the performance of this type of protocol, with proposals based on virtual photon subtraction~\cite{effort1,effort2}, unidimensional modulation~\cite{effort3}, or discrete modulation~\cite{effort4}. While these protocols offered an improvement in the range of the asymmetric configuration; in which the relay is positioned within close range of one of the parties, their applicability in the symmetric configuration, in which the relay is positioned equidistant between the parties, was very limited. Only Refs.~\cite{effort1, effort2} offered any improvement over the original CV MDI protocol in the symmetric configuration.

In this work, we begin to bridge the rate-distance gap between DV and CV MDI protocols. In particular, we aim to improve the distance over which a rate is attainable in the symmetric configuration. In this case, a secret key rate under the original CV MDI protocol and ideal conditions is only attainable at very short distances corresponding to a \SI{0.75}{dB} loss \cite{ottaviani2015continuous}. In order to extend this range, we employ a post-selection (PS) regime. PS describes the ability of the parties to select only instances of the protocol in which they have an advantage over the eavesdropper, given a prescriptive map of the contribution of the possible signals. By discarding any other instances, the secret key rate is always positive, and the parties can communicate securely up to a distance at which the key rate drops below a minimum usability threshold.

Post selection of a CV protocol was first introduced by Silberhorn et al.~\cite{silberhorn2002continuous} where it allowed a secret key to be constructed for losses exceeding the previous limit of \SI{3}{\dB}. Later, the technique was generalized to thermal loss channels~\cite{heid2006efficiency,heid2007security} and the concept has been demonstrated in the experimental setting~\cite{symul2007experimental, lance2005no}. In this work, we consider the post-selection of an MDI protocol which includes a measurement at an untrusted relay. We perform post-selection over the relay measurement outcome as well as Alice and Bob's variables while assuming that Eve employs a collective attack in which she targets both the Alice-relay and Bob-relay links.

The paper is structured as follows: we begin by outlining the protocol in detail and follow the evolution of the modes. We then derive the mutual information between the parties and the Holevo bound in order to quantify Eve's information. Using these quantities we can build the single-point rate, which serves as a prescriptive map for the parties to select the advantageous channel uses. Finally, we calculate the post selected secret key rate of the protocol.

\section{The protocol}

\begin{figure}[t]
\includegraphics[width=\linewidth]{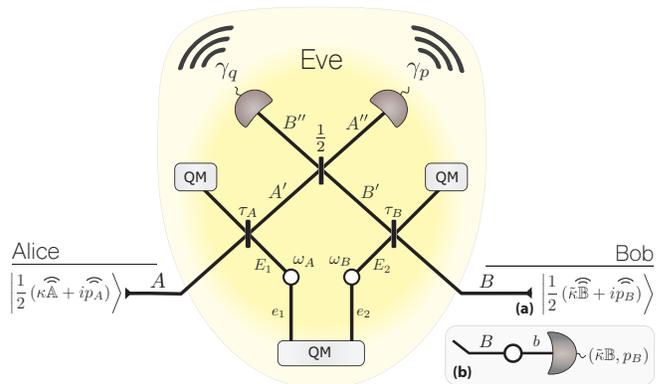}
\caption{Schematic of the protocol assuming the $q$-quadrature is chosen by the parties for reconciliation. (a): Alice and Bob send their coherent states to the relay. Eve is in possession of two two-mode squeezed vacuum states, denoted by white circles. She employs dual entangling cloner attacks, interacting with Alice and Bob's modes with beam splitters of transmissivity $\tau_A$ and $\tau_B$, respectively. The output modes $A'$ and $B'$ are mixed in a balanced beam splitter at the relay and the new output modes $A''$ and $B''$ are subsequently measured with homodyne $p$- and $q$-detection with corresponding outcomes $\gamma_p$ and $\gamma_q$, respectively that are publicly announced. After quantum communication ceases, Alice broadcasts $\absA$ and $p_A$ while Bob broadcasts $\absB$ and $p_B$. (b): In the restricted eavesdropping scenario Bob's action is modeled in the entanglement-based representation. He measures, with heterodyne detection, one mode $b$ of a two-mode squeezed vacuum state of variance $\mu$ obtaining the outcome $(\bobGuess\absB, p_B)$. This action prepares a coherent state in the conjugate mode $B$ that is subsequently sent to the relay.
}
\label{fig:CVMDI_PS_SCHEME}
\end{figure}

\begin{figure*}[t]
\includegraphics[width=0.75\textwidth]{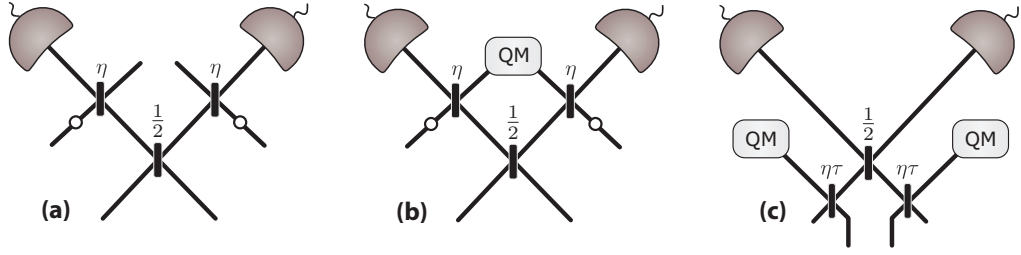}
\caption{Models of inefficiency in homodyne detection at the relay using beam splitters. (a) depicts a trusted noise scenario in which it is assumed that Eve does not have access to the output of the beam splitters. (b) assumes that the outputs of the beam splitters are added to Eve's quantum memory for later measurement. (c) depicts a simplification in the symmetric case ($\tau_A=\tau_B=\tau$) and with $S=1$ in which the transmissivities of the Alice-relay and Bob-relay links are scaled by a factor of $\eta$ to model the effect of beam splitters at detectors.}
\label{fig:experimentRelay}
\end{figure*}

Let us begin our analysis by outlining our protocol which is shown schematically in Fig.~\ref{fig:CVMDI_PS_SCHEME}. The secure parties that we label Alice and Bob are each connected to a relay with fiber optic links. We assume that both parties have access to a general Gaussian distribution of the form
\begin{equation}
	p(x, \sigma) = \frac{1}{\sqrt{2\pi \sigma}}\exp\left(-\frac{x^2}{2\sigma}\right).
\end{equation} 
In each use of the protocol, Alice draws two random numbers $q_A$ and $p_A$ from her Gaussian distribution with variance $\sigma_A$. From these two numbers, she extracts absolute values $|q_A|=\absA$ and $|p_A|=\absA'$ and signs $\kappa$ and $\kappa'$, respectively. For both $\kappa$ and $\kappa'$, she records bit values 0(1) if the sign is positive(negative). She proceeds to prepare a coherent state of the form $\left|\frac{1}{2}(\kappa \absA + i\kappa' \absA')\right\rangle$ and sends it to the relay via a quantum channel. Bob follows a similar procedure, generating two random numbers $q_B$ and $p_B$ using his Gaussian distribution with a generally different variance $\sigma_B$. He generates a state of the form $\left|\frac{1}{2}(\bobGuess \absB + i\bobGuess' \absB')\right\rangle$ and sends it to the relay.

After quantum communication ceases, the parties perform basis reconciliation. If the $q$-quadrature is chosen, the variables $\kappa'$ and $\bobGuess'$ are ignored. Alice publicly broadcasts $\absA$ and $p_A$ while Bob broadcasts $\absB$ and $p_B$ and attempts to reconcile his variable $\bobGuess$ with Alice's variable $\kappa$. Alternatively, if the $p$-quadrature is chosen, the relevant variables become $\kappa'$ and $\bobGuess'$. Alice broadcasts $\absA'$ and $q_A$ while Bob broadcasts $\absB'$ and $q_B$.

We assume that Eve employs dual entangling cloner attacks in which she inserts beam splitters of transmissivity $\tau_A$ and $\tau_B$ into lossless Alice-relay and Bob-relay channels, respectively. She uses the beam splitters to mix Alice's mode $A$ with her mode $E_1$ and Bob's mode $B$ with her mode $E_2$. The modes $E_1$ and $E_2$ each form one half of independent two-mode squeezed vacuum (TMSV) states with conjugate modes $e_1$ and $e_2$, and variances $\omega_A$ and $\omega_B$, respectively. She stores the outputs from one port of each beam splitter in a quantum memory and sends the remaining outputs $A'$ and $B'$ to the relay where they are mixed in a balanced beam splitter with outputs $A''$ and $B''$ that are subsequently measured with homodyne detection in the $p$- and $q$- quadratures, respectively. The corresponding outcomes $\gamma_p$ and $\gamma_q$ are publicly broadcast as $\boldsymbol{\gamma}=(\gamma_q,\gamma_p)$. 

To model detector inefficiencies, we can treat the modes $A''$ and $B''$ as passing through beam splitters of transmissivity $\eta$ where they are each mixed with one half of separate TMSV states with identical variance $S$ before arriving at 100\% efficient homodyne detectors. We may assume that the noise of the detectors is untrusted, in which case we assume the TMSV states are part of Eve's state and are included in the calculation of Eve's information, or trusted, in which case they are discarded. If $S=1$, and $\tau_A=\tau_B=\tau$, the detector inefficiencies can be modeled without considering beam splitter interactions at the relay by absorbing the detector efficiency parameter into the transmissivities of the links such that $\tau\to\eta\tau$. We outline each model schematically in Fig.~\ref{fig:experimentRelay}.

In this paper, our goal is to establish the post-selected asymptotic key rate of the protocol, $R_{\text{PS}}$. However, our initial objective is to obtain a formula for the standard asymptotic secret key rate $R$ which is given by the difference in the reconcilable information between the trusted parties, $\beta I_{AB}$ where $\beta$ is the reconciliation efficiency and $I_{AB}$ is the mutual information between the parties, and the Holevo bound $\chi$ which quantifies the maximum information Eve may attain about the secret variable depending on the particular attack,
\begin{equation}
	R = \beta I_{AB}-\chi.
\end{equation}
To this end, we follow the propagation of the covariance matrix (CM) of the total Alice-Bob-Eve system and its associated mean value. As each use of the protocol is Gaussian, these are the only tools we need to compute the probabilities and states needed to derive the key rate. After this step is complete, we explain the post-selection procedure which allows us to extend the range of the protocol.

The initial covariance matrix of the total system is given by
\begin{equation}
\mathbf{V}_{A B \eveStates|\kappa\bobGuess\kappa'\bobGuess'\absA\absB\absA'\absB'}=\mathbf{I}_A\oplus \mathbf{I}_B \oplus \mathbf{V}_{\eveStates},
\end{equation}
where $\mathbf{V}_{\eveStates}$ is Eve's initial CM, which, assuming she controls the detector noise at the relay, is given by
\begin{align}
	\mathbf{V}_{\eveStates} = &\mathbf{V}_{\text{TMSV}}(\omega_A)\oplus\mathbf{V}_{\text{TMSV}}(\omega_B)\nonumber \\
	&	\oplus \mathbf{V}_{\text{TMSV}}(S)\oplus\mathbf{V}_{\text{TMSV}}(S),
\end{align}
with $\mathbf{V}_{\text{TMSV}}(\mu)$ being the CM of a TMSV state with variance $\mu$ given by
\begin{equation}
	\mathbf{V}_{\text{TMSV}}(\mu) = \left(
	\begin{array}{cc}
		\mu \mathbf{I} & \sqrt{\mu^2-1}\mathbf{Z} \\
		\sqrt{\mu^2-1}\mathbf{Z} & \mu \mathbf{I}
	\end{array}
	\right),
	\label{InputCM}
\end{equation}
 where $\mathbf{Z}=\diag(1, -1)$ and $\mathbf{I}$ is the 2$\times$2 identity matrix. The mean value of the combined system of Alice and Bob is given by
\begin{equation}
  \bar{\mathbf{x}}_{AB|\kappa\bobGuess\kappa'\bobGuess'\absA\absB\absA'\absB'} = (\kappa\absA, \kappa'\absA', \bobGuess\absB,\bobGuess'\absB')\tran,
\end{equation}
while the mean value of Eve's system can be taken initially as zero. The action of all of the beam splitters can be encapsulated by a unitary operator $\hat{\mathbf{T}}$ that, when applied to the system, gives the post-propagation CM $\mathbf{V}_{A'' B'' \evePrime|\kappa\bobGuess\kappa'\bobGuess'\absA\absB\absA'\absB'}$ and mean value $\bar{\mathbf{x}}_{A'' B'' \evePrime|\kappa\bobGuess\kappa'\bobGuess'\absA\absB\absA'\absB'}$. Eve's CM with conditioning on $\boldsymbol{\gamma}$ is obtained by performing the homodyne measurements at the relay on the modes $A''$ and $B''$ in the $p$- and $q$-quadrature, respectively. The measurement outcome in the $q$-quadrature, $\gamma_q$ with conditioning on the measurement outcome of the $p$-quadrature, $\gamma_p$, is given by
\begin{align}
	&p(\gamma_q|\kappa\bobGuess\kappa'\bobGuess'\absA\absB \absA'\absB'\gamma_p) = \frac{1}{\sqrt{2\pi \relayVarOQ}}\times \nonumber \\
	&\times \exp\left[-\frac{1}{2 \relayVarOQ}\left(\gamma + \sqrt{\frac{\eta}{2}}(\kappa\absA\sqrt{\tau_A} - \bobGuess\absB\sqrt{\tau_B})\right)^2\right],
\end{align} 
and in the reverse case we have
\begin{align}
	&p(\gamma_p|\kappa\bobGuess\kappa'\bobGuess'\absA\absB \absA'\absB'\gamma_q) = \frac{1}{\sqrt{2\pi \relayVarOQ}}\times \nonumber \\
	&\times \exp\left[-\frac{1}{2 \relayVarOQ}\left(\gamma - \sqrt{\frac{\eta}{2}}(\kappa'\absA'\sqrt{\tau_A} + \bobGuess'\absB'\sqrt{\tau_B})\right)^2\right],
\end{align} 
where
\begin{equation}
\relayVarOQ = (1-\eta)S + \frac{\eta}{2}[\tau_A + \tau_B + (1-\tau_A)\omega_A + (1-\tau_B)\omega_B].
	\label{MDIPS1QCohVar}
\end{equation}
At this point, it is important to be aware that the measurement outcomes in the two quadratures are independent:
\begin{align}
	p(\gamma_q|\kappa\bobGuess\kappa'\bobGuess'\absA\absB \absA'\absB'\gamma_p)&=p(\gamma_q|\kappa\bobGuess\absA\absB) \\
	p(\gamma_p|\kappa\bobGuess\kappa'\bobGuess'\absA\absB \absA'\absB'\gamma_q)&=p(\gamma_p|\kappa'\bobGuess'\absA'\absB').
\end{align}  
This means that when the parties agree on their choice of quadrature in the basis reconciliation step, the remaining quadrature does not affect the rate. Moreover, the rate is independent of this choice. This fact allows us to calculate the rate using only one quadrature while ignoring the variables associated with its conjugate. Therefore, we will arbitrarily choose the $q$-quadrature for our forthcoming calculation of the rate and we will employ the refined notation $\gamma\equiv\gamma_q$ while ignoring the variables $\kappa'$, $\bobGuess'$, $\absA'$ and $\absB'$.  

\subsection{Restricted eavesdropping\label{restrictedProtocolIntro}}

If Bob broadcasts the tuple $(\absB,p_B)$ or $(q_B,\absB')$, he ensures that both parties can independently establish which instances of the protocol should be included in the final key. Such a communication step is likely a necessity in any post-selection protocol, however, there may be a more optimal strategy that reduces the amount of information Bob must broadcast and therefore the amount of information Eve gains. As an example, it may be possible for Bob to reveal the string of good instances at the end of the protocol as opposed to broadcasting his measurement data in each use. A strategy such as this would yield a secret key rate that lies in between the achievable lower bound in which Bob broadcasts the aforementioned tuples in every use of the protocol, and the upper bound in which no information is broadcast by Bob. An alternative way to think about the latter is to consider a \textit{restricted eavesdropping} scenario in which Eve does not make use of the information broadcast by Bob in her attack. In this context, it is possible to compute the upper bound on the secret key rate by computing Eve's states without conditioning on Bob's measurement outcome. To establish Eve's states in this case, we need to consider an entanglement-based version of the protocol as shown in Fig.~\ref{fig:CVMDI_PS_SCHEME}(b). Bob's action may be modeled as measuring one mode of a TMSV state with variance $\mu$. The amplitude of the coherent states $\ket{\tilde{\beta}}$ remotely prepared as a result of this process is related to the measurement outcome $\boldsymbol{\beta}$ by
\begin{equation}
	\tilde{\beta} = \xi\boldsymbol{\beta}^*,\quad \xi=\sqrt{\frac{\mu+1}{\mu-1}}.
\end{equation}
We label Bob's heterodyne measurement outcome $(\bobGuess \absB, \bobGuess'\absB')$.

For our analysis, we again consider only the $q$-quadrature using the fact that the quadratures are uncorrelated. After applying the beam splitter operation to the CM and mean value, we obtain the relay measurement outcome $\gamma\equiv\gamma_q$ with probability
\begin{equation}
  p(\gamma|\kappa\absA)=\frac{1}{\sqrt{2\pi \relayVarOQRE}} \exp\left[-\frac{1}{2 \relayVarOQRE}\left(\gamma+\kappa\absA\sqrt{\frac{1}{2}\eta\tau_A} \right)^2\right],
  \label{probGamma}
\end{equation}
where
\begin{equation}
	\relayVarOQRE=(1-\eta)S + \frac{\eta}{2}\left[\tau_A+\tau_B\mu + (1-\tau_A)\omega_A + (1-\tau_B)\omega_B\right].
\end{equation}
After the relay measurements, the CM and mean value of the remaining system become $\mathbf{V}_{b\evePrime|\kappa\absA\boldsymbol{\gamma}}$ and $\bar{\mathbf{x}}_{ b\evePrime |\kappa\absA\boldsymbol{\gamma}}$. Eve's CM and mean value are obtained by tracing out Bob's remaining mode $b$. In the final step, Bob performs a heterodyne measurement on his retained mode. The associated probability distribution $p(\bobGuess\absB, p_B|\kappa\absA\relayQ)$ and by integrating over $p_B$ we obtain
\begin{align}
&p(\bobGuess\absB|\kappa\absA\relayQ) =\frac{1}{\sqrt{2\pi V_b}}\times\nonumber \\
	&\exp\left[-\frac{1}{2 V_b}\left(\bobGuess\absB -\sqrt{(\mu^2-1)\frac{\eta\tau_B}{2}}\frac{\left(\relayQ + \kappa\absA\sqrt{\frac{1}{2}\eta\tau_A}\right)}{\relayVarOQ}\right)^2\right]
\end{align}
where
\begin{equation}
V_b = (\mu+1)\left(1-\frac{\mu-1}{\relayVarOQ}\frac{\eta\tau_B}{2}\right).
\end{equation}
In the following sections, we will derive the secret key rate of the protocol for both eavesdropping scenarios based on the secret encoding variable $\kappa$ and Bob's variable $\bobGuess$. We first compute the mutual information then the Holevo bound and, finally, we will introduce the post-selection procedure and calculate the post-selected rate.

\section{Mutual Information}
The first step in the calculations of the secret key rate is to establish the mutual information between Alice and Bob using the protocol outputs. The mutual information formula is given, independent of the eavesdropping strategy, by
\begin{equation}
	I(\kappa:\bobGuess |\absA\absB\relayQ) = H(\kappa|\absA\absB\relayQ) - H(\kappa|\bobGuess\absA\absB\relayQ),
\end{equation}
where, for random variables $X$ and $Y$, $H(X|Y) =\int p(y)H_{X|y}\,\mathrm{d}y$ is the conditional entropy of $X$ given $Y$ and $H_{X|y}$ is the entropy of $X$ conditioned on $Y$ taking the value $y$. The first term of the mutual information is therefore given by
\begin{equation}
	H(\kappa|\absA\absB\relayQ)=\int p(\absA\absB\relayQ)H_{\kappa|\absA\absB\relayQ}\,\mathrm{d}\absA\,\mathrm{d}\absB\,\mathrm{d}\relayQ,
\end{equation}
while the second may be expressed as
\begin{align}
	H(\kappa|&\bobGuess\absA\absB\relayQ)=\nonumber \\
	&\int p(\absA\absB\relayQ)\sum_{\bobGuess} p(\bobGuess|\absA\absB\relayQ)H_{\kappa|\bobGuess\absA\absB\relayQ}\,\mathrm{d}\absA\,\mathrm{d}\absB\,\mathrm{d}\relayQ,
\end{align}
where $H_{\kappa|\absA\absB\relayQ}$ and $H_{\kappa|\bobGuess\absA\absB\relayQ}$ reduce to the binary entropy of respective probabilities $p(\kappa|\absA\absB\relayQ)$ and $p(\kappa|\bobGuess\absA\absB\relayQ)$. We can derive the latter probability using Bayes' theorem as
\begin{align}
	p(&\kappa |\bobGuess\absA\absB\gamma) = \frac{p(\relayQ|\kappa\bobGuess\absA\absB)p(\kappa|\bobGuess\absA\absB)}{\sum_\kappa p(\relayQ|\kappa\bobGuess\absA\absB)p(\kappa|\bobGuess\absA\absB)}\nonumber \\
	&=\frac{1}{1+\exp\left[2\kappa\absA\sqrt{\frac{1}{2}\eta\tau_A}\left(\gamma - \bobGuess\absB\sqrt{\frac{1}{2}\eta\tau_B}\right)\relayVarOQ^{-1}\right]},
\end{align}
where we have used the fact that $\kappa$, $\bobGuess$, $\absA$ and $\absB$ are independent variables. Using the same logic, we derive
\begin{align}
	p(&\bobGuess|\kappa\absA\absB\gamma) =\nonumber\\
	&\frac{1}{1+\exp\left[-2\bobGuess\absB\sqrt{\frac{1}{2}\eta\tau_B}\left(\gamma + \kappa\absA\sqrt{\frac{1}{2}\eta\tau_A}\right)\relayVarOQ^{-1}\right]}.
\end{align}
We also require the probabilities of each of $\kappa$ and $\bobGuess$ with conditioning on $\absA$, $\absB$ and $\gamma$ only. We have
\begin{align}
	&p(\kappa|\absA\absB\gamma) = \frac{\sum_{\bobGuess} p(\relayQ|\kappa\bobGuess\absA\absB)p(\kappa\bobGuess|\absA\absB)}{\sum_{\kappa,\bobGuess} p(\relayQ|\kappa\bobGuess\absA\absB)p(\kappa\bobGuess|\absA\absB)}=\nonumber\\
	&\frac{1}{1+\left(\frac{p(+|1\absA\absB\gamma)}{p(-|0\absA\absB\gamma)}\right)^{\kappa} \exp\left[2\kappa\sqrt{\frac{1}{2}\eta}(\absB\sqrt{\tau_B}+\absA\sqrt{\tau_A})\relayVarOQ^{-1}\right]},
\end{align}
where we note that $p(\kappa\bobGuess|\absA\absB)=1/4$ for all combinations of $\kappa$ and $\bobGuess$ due to the independence of the variables. Using the same logic we obtain the remaining probability required for the calculation of the conditional entropies,
\begin{align}
	&p(\bobGuess|\absA\absB\gamma)=\\
	&\frac{1}{1+\left(\frac{p(0|-\absA\absB\gamma)}{p(1|+\absA\absB\gamma)}\right)^{\bobGuess} \exp\left[-2\bobGuess\sqrt{\frac{1}{2}\eta}(\absB\sqrt{\tau_B}+\absA\sqrt{\tau_A})\relayVarOQ^{-1}\right]}.
\end{align}
The final probability we require is the total probability of all of the post-selection variables which is given by
\begin{equation}
	p(\absA\absB\relayQ) = \sum_{\kappa,\bobGuess}p(\relayQ|\kappa\bobGuess\absA\absB)p(\kappa\absA)p(\bobGuess\absB).
\end{equation}

The probabilities for the computation of the mutual information in the restricted eavesdropping scenario are slightly more complicated due to Bob's TMSV state, however, the first conditional probability is easily attainable as
\begin{align}
p&(\bobGuess|\kappa\absA\absB\relayQ) =\frac{p(\bobGuess\absB|\kappa\absA\relayQ)}{\sum{p(\bobGuess\absB|\kappa\absA\relayQ)}} \nonumber\\
	&=\frac{1}{1+\exp\left[-2\bobGuess\absB\left(\relayQ + \kappa\absA\sqrt{\frac{1}{2}\eta\tau_A}\right)\Delta\ \relayVarOQRE'^{-1}\right]},
\end{align}
where we have defined
\begin{equation}
	\relayVarOQRE'=(1-\eta)S + \frac{\eta}{2}\left[\tau_A+\tau_B + \omega_A(1-\tau_A) + \omega_B(1-\tau_B)\right]
\end{equation}
and
\begin{equation}
	\Delta = \sqrt{\frac{\eta}{2}\frac{1}{\tau_B}}\sqrt{\frac{\mu-1}{\mu+1}}.
\end{equation}
In order to calculate the reverse probability $p(\kappa|\bobGuess\absA\absB\relayQ)$, we first compute
\begin{align}
	p(\kappa|\absA\gamma) &= \frac{p(\gamma|\kappa\absA)}{\sum_\kappa p(\gamma|\kappa\absA)}\\
	&=\frac{1}{1+\exp\left(2\kappa\absA\gamma\sqrt{\frac{1}{2}\eta\tau_A}\relayVarOQRE^{-1}\right)},
\end{align}
then the required probability can be derived as
\begin{align}
p(\kappa|\bobGuess &\absA\absB\relayQ) = \frac{ p(\bobGuess\absB|\kappa\absA\relayQ)p(\kappa|\absA\relayQ)}{\sum_{\kappa}p(\bobGuess\absB|\kappa\absA\relayQ)p(\kappa|\absA\relayQ)} \nonumber\\
	=\ &\frac{1}{1+\exp\left[2\kappa\absA\sqrt{\frac{1}{2}\eta\tau_A}\left(\gamma' - \bobGuess\absB\Delta\right)\relayVarOQRE'^{-1}\right]}
\end{align}
where we have defined
\begin{equation}
	\gamma' = \frac{1}{\relayVarOQRE}\left(\relayVarOQRE'+\frac{\eta}{2}\frac{1}{\tau_B}(\mu-1)\right)\gamma.
\end{equation}

We can now compute the total probabilities of $\kappa$ and $\bobGuess$ as
\begin{align}
p(\kappa|&\absA\absB\relayQ) = \frac{\sum_{\bobGuess} p(\bobGuess\absB|\kappa\absA\relayQ)p(\kappa|\absA\relayQ)}{\sum_{\kappa,\bobGuess}p(\bobGuess\absB|\kappa\absA\relayQ)p(\kappa|\absA\relayQ)}\nonumber\\
	&=\frac{1}{1+\Xi_\kappa\exp\left[2\kappa\absA\sqrt{\frac{1}{2}\eta\tau_A}\left(\gamma' +\absB\Delta\right)\relayVarOQRE'^{-1}\right]}
\end{align}
and
\begin{align}
p(\bobGuess|&\absA\absB\relayQ)=\frac{1}{1+\Xi_{\bobGuess}\exp\left[-2\bobGuess\absB\left(\relayQ - \absA\sqrt{\frac{1}{2}\eta\tau_A}\right)\relayVarOQRE'^{-1}\Delta\right]}
\end{align}
with 
\begin{equation}
	\Xi_{m} = \left(\frac{p(1|+\absA\absB\relayQ)}{p(1|-\absA\absB\relayQ)}\right)^{m}.
\end{equation}
Finally, the total probability of the three post-selection variables becomes
\begin{equation}
	p(\absA\absB\relayQ) = \sum_{\kappa,\bobGuess}p(\bobGuess\absB|\kappa\absA\gamma)p(\gamma|\kappa\absA)p(\kappa\absA).
\end{equation}


\section{Holevo bound}
In our consideration of Eve's accessible information on the secret variable, we use the Holevo bound, which quantifies the maximum amount of information Eve may attain using any strategy permitted by the laws of quantum mechanics. We may write the bound as 
\begin{equation}
\chi(\eveStatesPrime:\kappa|\absA\absB\relayQ) = S(\eveStatesPrime|\absA\absB\relayQ) - S(\eveStatesPrime|\kappa\absA\absB\relayQ),
\end{equation}
where $S(X|x) :=\int p(x)S(\hat{\rho}_{X|x})\,\mathrm{d}x$ is the conditional von Neumann entropy (VNE) of system $X$ on variable $x$ with corresponding probability distribution $p(x)$, and $S(\hat{\rho})$ is the VNE of state $\hat{\rho}$, defined as
\begin{equation}
	S(\hat{\rho}) = -\sum_i \lambda_i\log_2\lambda_i,
	\label{StateVNE}
\end{equation}
where $\{\lambda_i\}$ are the eigenvalues of $\hat{\rho}$. The first term of the Holevo bound can be written as
\begin{equation}
	S(\eveStatesPrime|\absA\absB\relayQ) = \int p(\absA\absB\relayQ)S(\hat{\rho}_{\eveStatesPrime|\absA\absB\relayQ})\, \mathrm{d}\absA\,\mathrm{d}\absB\,\mathrm{d}\relayQ,
	\label{HBT1TB}
\end{equation}
where $\hat{\rho}_{\eveStatesPrime|\absA\absB\relayQ}$ is Eve's total state, which can be derived from the output state after propagation as
\begin{equation}
	\hat{\rho}_{\eveStatesPrime|\absA\absB\relayQ}=\sum_{\kappa,\bobGuess}p(\kappa\bobGuess|\absA\absB\relayQ)\hat{\rho}_{\eveStatesPrime|\kappa\bobGuess\absA\absB\relayQ}.
	\label{EveTotalState2}
\end{equation} 
Similarly, the second (conditional) term is given by
\begin{align}
	&S(\eveStatesPrime|\kappa\absA\absB\gamma)=\nonumber\\
	&\int p(\absA\absB\relayQ) \sum_\kappa p(\kappa|\absA\absB\relayQ)S\left(\hat{\rho}_{\eveStatesPrime|\kappa\absA\absB\relayQ}\right)\,\mathrm{d}\absA\,\mathrm{d}\absB\,\mathrm{d}\relayQ,
		\label{EveConditionalState2}
\end{align}
where $\hat{\rho}_{\eveStatesPrime|\kappa\absA\absB\relayQ}$ is Eve's conditional state given by
\begin{equation}
	\hat{\rho}_{\eveStatesPrime|\kappa\absA\absB\relayQ}=\sum_{\bobGuess}p(\bobGuess|\kappa\absA\absB\relayQ)\hat{\rho}_{\eveStatesPrime|\kappa\bobGuess\absA\absB\relayQ}.
\end{equation}
Note that Eve's total and conditional states in addition to the output state of the protocol are implicitly conditioned on the variables associated with the $p$-quadrature, but as they do not impact the rate, they need not be included in the calculation explicitly. However, it is important to note that the relay operation includes a measurement of the $p$-quadrature, which guarantees that the state of Eve's system after propagation, namely $\hat{\rho}_{\eveStatesPrime|\kappa\bobGuess\absA\absB\relayQ}$, is a pure state.

Neither the total nor the condition states of Eve's system are Gaussian, and computing their entropy directly in the Fock basis is a difficult problem. Instead, we follow a method originating from Refs \cite{heid2006efficiency, heid2007security} for one-way protocols with coherent states, and with little added complexity we derive the equivalent method for the MDI protocol with coherent states. Because Eve's state emerging from the protocol is pure, it can be written in the shorthand
notation
\begin{equation}
	\hat{\rho}_{\eveStatesPrime|\kappa\bobGuess\absA\absB\relayQ} = \hat{\mathfrak{E}}'^{\absA\absB\relayQ}_{\kappa\bobGuess} = \left|{\eveStatesPrime^{\absA\absB\relayQ}_{\kappa\bobGuess}}\right\rangle\left\langle{\eveStatesPrime^{\absA\absB\relayQ}_{\kappa\bobGuess}}\right|.
\end{equation}
For convenience, we also introduce the shorthand notation
\begin{align}
	& p_{\kappa\bobGuess}^{\absA\absB\relayQ}\equiv p(\kappa\bobGuess|\absA\absB\relayQ) \\
\text{and} \quad & p_{\bobGuess|\kappa}^{\absA\absB\relayQ}\equiv p(\bobGuess|\kappa\absA\absB\relayQ).
\end{align}
Using the broadcast values $\absA$, $p_A$, $\absB$, $p_B$ and $\boldsymbol{\gamma}$, Eve knows that her total state is a convex combination of the four states $\left|{\eveStatesPrime^{\absA\absB\relayQ}_{0+}}\right\rangle$,$\left|{\eveStatesPrime^{\absA\absB\relayQ}_{0-}}\right\rangle$,$\left|{\eveStatesPrime^{\absA\absB\relayQ}_{1+}}\right\rangle$ and $\left|{\eveStatesPrime^{\absA\absB\relayQ}_{1-}}\right\rangle$ and her state can be expressed in a four-dimensional space. Note that in our notation we use Alice's assigned bit values 0(1) to represent $\kappa=+(-)$ in order to aide distinguishability between $\kappa$ and $\bobGuess$. 

Let us re-write the total state in Eq.~(\ref{EveTotalState2}) as
\begin{equation}
	\hat{\rho}_{\eveStatesPrime|\absA\absB\relayQ} = \sum_{\kappa,\bobGuess}p_{\kappa\bobGuess}^{\absA\absB\relayQ}\left|{\eveStatesPrime^{\absA\absB\relayQ}_{\kappa\bobGuess}}\right\rangle\left\langle{\eveStatesPrime^{\absA\absB\relayQ}_{\kappa\bobGuess}}\right|.
	\label{EveTotalStateSimp}
\end{equation}

To examine the information held by Eve in her state we can compute the matrix of all overlaps $\mathbf{S}$, whose elements $S_{ij}$ are given by the overlaps $\left\langle\eveStatesPrime^{\absA\absB\relayQ}_{\kappa_1\bobGuess_1}\middle|\eveStatesPrime^{\absA\absB\relayQ}_{\kappa_2\bobGuess_2}\right\rangle$ of Eve's possible states. We may write the matrix of all overlaps as
\begin{equation}
\mathbf{S}=\ \ \begin{blockarray}{ccccc}
0+ & 0- & 1+ & 1- & \\
\begin{block}{(cccc)c}
 1 & B & A & AB & \ 0+ \\
 B & 1 & AB & A & \ 0- \\
 A & AB & 1 & B & \ 1+ \\
 AB & A & B & 1 & \ 1- \\
\end{block}
\end{blockarray}
 \end{equation}
where we have ignored phase factors that may always be removed by multiplying the states $\left|\eveStatesPrime^{\absA\absB\relayQ}_{\kappa\bobGuess}\right\rangle$ by other appropriate phase factors. The matrix of overlaps reveals the inter-relationship between the basis vectors in Eve's total state. It can be seen that the matrix is expressible in tensor-product form as
\begin{equation}
	\mathbf{S}=\begin{pmatrix}
  1 & A \\
  A & 1
\end{pmatrix}
\otimes
\begin{pmatrix}
  1 & B \\
  B & 1
\end{pmatrix},
\end{equation}
which implies that Eve's state is the product of two states in two-dimensional Hilbert spaces, which we write as
\begin{equation}
	\left|\eveStatesPrime^{\absA\absB\relayQ}_{\kappa\bobGuess}\right\rangle = \left|\eveStatesPrime^{\absA\absB\relayQ}_{\kappa}\right\rangle\left|\eveStatesPrime^{\absA\absB\relayQ}_{\bobGuess}\right\rangle.
\end{equation}
The individual states can be expanded as 
\begin{align}
\left|\eveStatesPrime^{\absA\absB\relayQ}_{0}\right\rangle &= c_0\ket{\Phi_0} + c_1\ket{\Phi_1}\\
\left|\eveStatesPrime^{\absA\absB\relayQ}_{1}\right\rangle &= c_0\ket{\Phi_0} - c_1\ket{\Phi_1},
\end{align}
and
\begin{align}
\left|\eveStatesPrime^{\absA\absB\relayQ}_{+}\right\rangle &= c_+\ket{\Phi_+} + c_-\ket{\Phi_-}\\
\left|\eveStatesPrime^{\absA\absB\relayQ}_{-}\right\rangle&= c_+\ket{\Phi_+} - c_-\ket{\Phi_-},
\end{align}
where $\{\ket{\Phi_0},\ket{\Phi_1}\}$ and $\{\ket{\Phi_+},\ket{\Phi_-}\}$ are orthonormal basis sets for the Hilbert spaces spanned by $\left|\eveStatesPrime^{\absA\absB\relayQ}_{\kappa}\right\rangle$ and $\left|\eveStatesPrime^{\absA\absB\relayQ}_{\bobGuess}\right\rangle$, respectively.

Our focus now turns to relating the coefficients to the overlaps $A$ and $B$. We perform the following inner products
\begin{align}
	\left\langle\eveStatesPrime^{\absA\absB\relayQ}_{0}\middle|\eveStatesPrime^{\absA\absB\relayQ}_{0}\right\rangle &= |c_0|^2 + |c_1|^2 = 1\\
	\left\langle\eveStatesPrime^{\absA\absB\relayQ}_{0}\middle|\eveStatesPrime^{\absA\absB\relayQ}_{1}\right\rangle &= |c_0|^2-|c_1|^2 	= A,
\end{align}
from which we obtain expressions for the absolute values of the coefficients $c_0$ and $c_1$ of
\begin{align}
	&|c_0|^2 = \frac{1}{2}\left(1+A\right)\\
	\text{and}\quad &|c_1|^2 = \frac{1}{2}\left(1-A\right),
\end{align}
and following a similar calculation we arrive at the following expressions for the absolute values of the remaining coefficients
\begin{align}
	&|c_+|^2 = \frac{1}{2}\left(1+B\right)\\
	\text{and}\quad &|c_-|^2 = \frac{1}{2}\left(1-B\right).
\end{align}

The values $A$ and $B$ are computed from the overlap formula for Gaussian states \cite{banchi2015quantum}, which, for two pure states $\hat{\rho}_1$ and $\hat{\rho}_2$ with the same CM, $\mathbf{V}$ and different mean values $\bar{\mathbf{x}}_1$ and $\bar{\mathbf{x}}_2$, reduces to
\begin{equation}
	\tr(\hat{\rho}_1\hat{\rho}_2) = \exp\left[-\frac{1}{4}(\bar{\mathbf{x}}_1-\bar{\mathbf{x}}_2)\tran \mathbf{V}^{-1}(\bar{\mathbf{x}}_1-\bar{\mathbf{x}}_2)\right],
	\label{FidSameCM}
\end{equation}
and our coefficients $A$ and $B$ become
\begin{equation}	
	A = \left\langle\eveStatesPrime^{\absA\absB\relayQ}_{0} \middle|\eveStatesPrime^{\absA\absB\relayQ}_{1}\right\rangle=\exp\left[-\frac{1}{2}\absA^2\left(1-\frac{\eta\tau_A}{\relayVarOQ}\right)\right],
\end{equation}
and 
\begin{equation}
	B = \left\langle\eveStatesPrime^{\absA\absB\relayQ}_{+} \middle|\eveStatesPrime^{\absA\absB\relayQ}_{-}\right\rangle=\exp\left[-\frac{1}{2}\absB^2\left(1-\frac{\eta\tau_B}{\relayVarOQ}\right)\right].
\end{equation}

We now have all of the tools required to compute Eve's total state using Eq.~(\ref{EveTotalStateSimp}). We arrive at the following matrix
\begin{widetext}
\begin{equation}
	\hat{\mathfrak{E}}'^{\absA\absB\relayQ}=
	\begin{pmatrix}
  |c_0|^2|c_+|^2 & |c_0|^2c_+c_-^* \Lambda(+,-,+,-) & |c_+|^2c_0c_1^*\Lambda(+,+,-,-) & c_0c_+c_1^*c_-^*\Lambda(+,-,-,+) \\
  |c_0|^2c_-c_+^*\Lambda(+,-,+,-) & |c_0|^2|c_-|^2 & c_0c_-c_1^*c_+^*\Lambda(+,-,-,+) & |c_-|^2c_0c_1^*\Lambda(+,+,-,-) \\
  |c_+|^2c_1c_0^*\Lambda(+,+,-,-) & c_1c_+c_0^*c_-^*\Lambda(+,-,-,+) & |c_1|^2|c_+|^2 & |c_1|^2c_+c_-^*\Lambda(+,-,+,-) \\
  c_1c_-c_0^*c_+^*\Lambda(+,-,-,+) & |c_-|^2c_1c_0^*\Lambda(+,+,-,-) & |c_1|^2c_0c_+^*\Lambda(+,-,+,-) & |c_1|^2|c_-|^2
	\end{pmatrix},
	\label{EveCMTOM}
\end{equation}
\end{widetext}
where we have defined
\begin{align}
	\Lambda(s_1, s_2, s_3, s_4) =&\ s_1p_{0+}^{\absA\absB\relayQ} + s_2p_{0-}^{\absA\absB\relayQ}+s_3p_{1+}^{\absA\absB\relayQ}+ s_4p_{1-}^{\absA\absB\relayQ}.
\end{align}
To obtain the entropy of the total state, we first compute the eigenvalues of Eq.~(\ref{EveCMTOM}) which amounts to solving a quartic equation in which the coefficients are combinations of the absolute values of the basis coefficients. We then compute their VNE using Eq.~(\ref{StateVNE}). This entropy is then substituted into Eq.~(\ref{HBT1TB}) to obtain the first term of the Holevo bound.

In order to compute the conditional state and the second term of the Holevo bound, we construct the density matrices of the conditional states. Firstly, we have
\begin{align}
	\hat{\mathfrak{E}}'^{\absA\absB\relayQ}_{0} =& \left|{\eveStatesPrime^{\absA\absB\relayQ}_{0}}\right\rangle\left\langle{\eveStatesPrime^{\absA\absB\relayQ}_{0}}\right|\otimes \bigg(p_{+|0}^{\absA\absB\relayQ}\left|{\eveStatesPrime^{\absA\absB\relayQ}_{+}}\right\rangle\left\langle{\eveStatesPrime^{\absA\absB\relayQ}_{+}}\right|\nonumber \\
	&+p_{-|0}^{\absA\absB\relayQ}\left|{\eveStatesPrime^{\absA\absB\relayQ}_{-}}\right\rangle\left\langle{\eveStatesPrime^{\absA\absB\relayQ}_{-}}\right|	\bigg),
\end{align}
which has corresponding eigenvalues
\begin{equation}
	\lambda_{1,2}^0 = \frac{1}{2}\left(1\pm\sqrt{1-16p_{+|0}^{\absA\absB\relayQ}p_{-|0}^{\absA\absB\relayQ}|c_-|^2|c_+|^2}\right).
\end{equation}
Then, for the counterpart state we have
\begin{align}
	\hat{\mathfrak{E}}'^{\absA\absB\relayQ}_{1} &= \left|{\eveStatesPrime^{\absA\absB\relayQ}_{1}}\right\rangle\left\langle{\eveStatesPrime^{\absA\absB\relayQ}_{1}}\right|\otimes \bigg(p_{+|1}^{\absA\absB\relayQ}\left|{\eveStatesPrime^{\absA\absB\relayQ}_{+}}\right\rangle\left\langle{\eveStatesPrime^{\absA\absB\relayQ}_{+}}\right|\nonumber \\
	&+p_{-|1}^{\absA\absB\relayQ}\left|{\eveStatesPrime^{\absA\absB\relayQ}_{-}}\right\rangle\left\langle{\eveStatesPrime^{\absA\absB\relayQ}_{-}}\right|	\bigg),
\end{align}
with eigenvalues
\begin{equation}
	\lambda_{1,2}^1 = \frac{1}{2}\left(1\pm\sqrt{1-16p_{+|1}^{\absA\absB\relayQ}p_{-|1}^{\absA\absB\relayQ}|c_-|^2|c_+|^2}\right).
\end{equation}
Using the eigenvalues of the two states, it is straightforward to compute the second term of the Holevo bound using Eq.~(\ref{EveConditionalState2}). 

\subsection{Restricted eavesdropping}
Let us now consider Eve's accessible information in the restricted eavesdropping scenario. In this case, Eve has to distinguish between two states corresponding to the two possible values of $\kappa$. Under these conditions, it is possible to consider both individual and collective attacks as we will outline in the following sections.

\subsubsection{Individual attacks}
Let us first examine the case in which Eve employs individual attacks, and may not access a quantum memory. In this case the mutual information between Alice and Eve, $I_{AE}$, can be estimated by from Eve's error probability using the fidelity, $F$ of Eve's two possible states, $\hat{\rho}_{\eveStatesPrime|+\absA\gamma}$ and $\hat{\rho}_{\eveStatesPrime|-\absA\gamma}$ which we compute using Eq.~(\ref{FidSameCM}). We apply the following lower bound \begin{equation}
F_-=\frac{1-\sqrt{1-F}}{2}
\end{equation}
in order to bound Eve's error probability from below, modeling a worst-case scenario for Alice and Bob \cite{pirandola2008computable}. The total expression for the mutual information $I_{AB}$ becomes
\begin{equation}
  I_{AE} =\int p(\absA\gamma)\left[1-H_2(F_{-})\right] \, \mathrm{d}\absA\,\mathrm{d}\gamma.
  \label{eq:individualIAE}
\end{equation}
where $H_2(p)$ is the binary entropy.

\subsubsection{Collective attacks}
In the case of collective attacks we must compute the Holevo bound in order to establish an upper-bound Eve's accessible information. The Holevo bound is given by
\begin{equation}
	\HBRE(\eveStatesPrime:\kappa|\absA\relayQ) = S(\eveStatesPrime|\absA\relayQ) - S(\eveStatesPrime|\kappa\absA\relayQ),
\end{equation}
where the first term can be written as
\begin{equation}
		S(\eveStatesPrime|\absA\relayQ) = \int p(\absA\relayQ)S(\hat{\rho}_{\eveStatesPrime|\absA\relayQ})\,\mathrm{d}\absA\,\mathrm{d}\relayQ,
		\label{eq:pureLossHBT1}
\end{equation}
where $\hat{\rho}_{\eveStatesPrime|\absA\relayQ}$ is the total state, given by
\begin{equation}
	\hat{\rho}_{\eveStatesPrime|\absA\gamma} = \sum_{\kappa}p(\kappa|\absA\gamma)\hat{\rho}_{\eveStatesPrime|\kappa\absA\gamma}.
	\label{totalStateProtocol}
\end{equation}
As it is derived from the sum of two Gaussian states, the total state is non-Gaussian. To avoid the difficulty in obtaining the entropy of this state from its photon statistics, we may employ a non-Gaussian entropy approximation which we derive in Appendix \ref{GEA}. Using the main result we may write the CM of the total state as
\begin{equation}
	\mathbf{V}_{\eveStatesPrime|\absA} = \mathbf{V}_{\eveStatesPrime|\kappa\absA} + p(+|\absA\relayQ)p(-|\absA\relayQ)\Delta\bar{\mathbf{x}}_{\eveStatesPrime}\cdot\Delta\bar{\mathbf{x}}\tran_{\eveStatesPrime},
\end{equation}
where $	\Delta\bar{\mathbf{x}}_{\eveStatesPrime} = \bar{\mathbf{x}}_{\eveStatesPrime|+\absA\relayQ} - \bar{\mathbf{x}}_{\eveStatesPrime|-\absA\relayQ}$. Taking the entropy of this state via the symplectic eigenvalues, $\{\nu_i\}$ of its CM provides an upper bound on the exact entropy of Eve's total state as it assumes this state to be Gaussian. We therefore have
\begin{equation}
	S(\hat{\rho}_{\kappa\absA\relayQ}) \leq S(\mathbf{V}_{\eveStatesPrime|\kappa\absA\relayQ}) = \sum_i h(\nu_i),
\end{equation}
where
\begin{equation}
	h(\nu) = \frac{\nu+1}{2}\log_2\frac{\nu+1}{2} - \frac{\nu-1}{2}\log_2\frac{\nu-1}{2}.
	\label{symplecticEvalsEnt}
\end{equation}

Meanwhile, the second term of the Holevo bound involves a Gaussian state and can be computed directly from the protocol output, independent of any measurement outcome. As described in Section~\ref{restrictedProtocolIntro}, Eve's CM $\mathbf{V}_{\eveStatesPrime|\kappa\absA}$ after the relay measurements is obtained by tracing out Bob's remaining mode. The entropy is then computed from the symplectic eigenvalues, $\{\upsilon_i\}$ of the remaining CM by
\begin{equation}
	S(\hat{\rho}_{\eveStatesPrime|\kappa\absA}) = S(\mathbf{V}_{\eveStatesPrime|\kappa\absA}) = \sum_i h(\upsilon_i),
\end{equation}
and the Holevo bound is reduced to the following expression
\begin{align}
	\chi(\eveStatesPrime:\kappa|\absA\relayQ) \leq \int p(\absA\relayQ)& S(\mathbf{V}_{\eveStatesPrime|\absA\relayQ})\, \mathrm{d}\absA\,\mathrm{d}\relayQ \nonumber\\
	&- S(\mathbf{V}_{\eveStatesPrime|\kappa\absA\relayQ}).
\end{align}

\section{Application of post-selection}
We have now computed all of the components required for the calculation of the secret key rate and we can now describe the post-selection step that improves the range of our protocol. Let us first write the mutual information as a single integrand in the following form
\begin{equation}
  I_{AB} = \int p(\absA\absB\relayQ) \tilde{I}_{AB}(\absA, \absB,\gamma) \, \mathrm{d}\absA\, \mathrm{d}\absB\, \mathrm{d}\relayQ,
\end{equation}
where we defined the \textit{single-point} mutual information $\tilde{I}_{\text{AB}}(\absA,\absB,\gamma) = H_{\kappa|\absA\absB\relayQ} - \sum_{\bobGuess} p(\bobGuess|\absA\absB\relayQ)H_{\kappa|\bobGuess\absA\absB\relayQ}$. Similarly, we can write the Holevo bound as a single integrand
\begin{equation}
	\chi = \int p(\absA\absB\gamma)\tilde{\chi}(\absA,\absB,\gamma)\, \mathrm{d}\absA\,\mathrm{d}\absB\,\mathrm{d}\gamma,
\end{equation}
with $\tilde{\chi}$ being the single-point Holevo bound given by 
\begin{align}
	\tilde{\chi} = S(\hat{\rho}_{\eveStatesPrime|\absA\absB\relayQ}) - \sum_\kappa p(\kappa|\absA\absB\relayQ)S(\rho_{\eveStatesPrime|\kappa\absA\absB\relayQ}).
\end{align}
In the same way, we define the following single-point Holevo bound for restricted eavesdropping, $\tilde{\chi}^\text{RE}$ for collective attacks and the single-point mutual information between Alice and Eve, $\tilde{I}_{AE}$ for individual attacks,
\begin{align}
	\HBRE &\leq S(\mathbf{V}_{\eveStatesPrime|\absA\relayQ}) - S(\mathbf{V}_{\eveStatesPrime|\kappa\absA\relayQ})\\
	\tilde{I}_{AE} &= 1-H_2(F_-).
\end{align}
Using these definitions, we may define the single-point rate, 
$\tilde{R}= \tilde{I}_{\text{AB}}-\tilde{\chi}$ for complete eavesdropping, $\tilde{R}= \tilde{I}_{\text{AB}}-\tilde{\chi}^{RE}$ for restricted eavesdropping and $\tilde{R}= \tilde{I}_{\text{AB}}-\tilde{I}_{\text{AE}}$ for individual attacks. We can then express the secret key rate in terms of the single-point rate as
\begin{equation}
	R = \int p(\absA\absB\relayQ) \tilde{R}(\absA, \absB,\relayQ) \, \mathrm{d}\absA\,\mathrm{d}\absB\,\mathrm{d}\relayQ.
\end{equation}

For post-selection, we are interested in the region where the single-point rate is positive so that the parties can choose to only include instances of the protocol that contribute positively to the key rate. We can therefore define the post-selected key rate as
\begin{equation}
	R_{\text{PS}} = \int p(\absA\absB\relayQ) \max\{\tilde{R}(\absA, \absB,\relayQ),0\} \, \mathrm{d}\absA \,\mathrm{d}\absB\,\mathrm{d}\relayQ.
\end{equation}
We can also define the post-selection area $\Gamma$, which is simply the region of the $\absA$-$\absB$-$\gamma$ volume in which the single-point rate is positive. Computing the post-selected rate amounts to integrating the single-point rate in this volume,
\begin{equation}
	R_{\text{PS}} = \int_\Gamma p(\absA\absB\relayQ) \tilde{R}(\absA, \absB,\relayQ) \, \mathrm{d}\absA \,\mathrm{d}\absB\,\mathrm{d}\relayQ.
\end{equation}

\section{Results}

\begin{figure}[t!]
	\includegraphics[width=\linewidth]{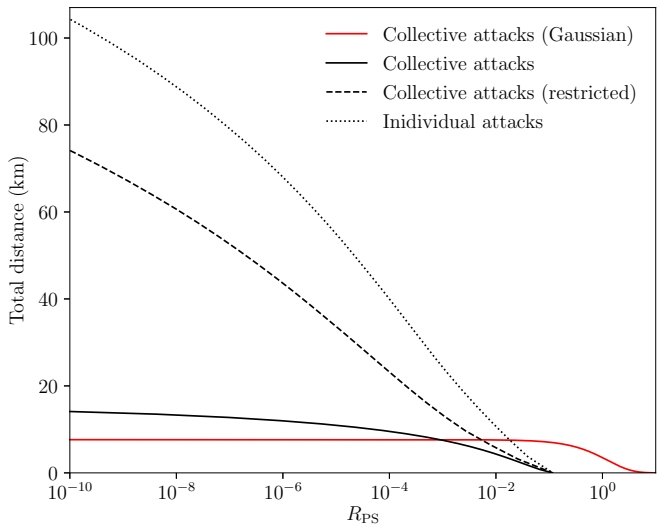}
	\caption{Rates of the pure-loss symmetric protocol as a function of the total distance between Alice and Bob with $\sigma_A$, $\sigma_B$ and $\mu$ optimized. The red line represents the rate of the symmetric Gaussian MDI protocol.}
	\label{fig:RatesBoth}
\end{figure}

Let us now present numerical results for the rates of our protocol under a variety of parameters. In order to express the rates as a function of the distance between the parties, we have first used the relation $\tau=10^{-\SI{}{\dB}/10}$ to express the transmissivity in terms of the loss in \SI{}{\dB}. Then, if the protocol is performed with standard optical fiber, the length of the links can be expressed in kilometers assuming a loss per kilometer of \SI{0.2}{\dB}/\SI{}{\km}. We use the excess noise to express the variances $\omega_A$ and $\omega_B$ in terms of the transmissivities of the channels. By considering each link to be a point-to-point channel we write
\begin{equation}
	\omega_{A(B)} = 1 + \epsilon_{A(B)}\frac{\eta\tau_{A(B)}/2}{1-\eta\tau_{A(B)}/2},
\end{equation}
where $\epsilon_{A(B)}$ is the excess noise in the Alice-relay (Bob-relay) links.

\begin{figure}[t!]
	\includegraphics[width=\linewidth]{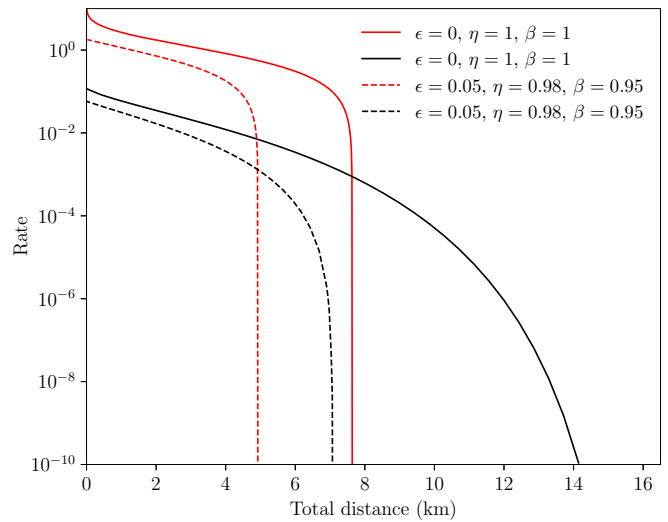}
	\caption{Rates of the symmetric protocol function of the total distance between Alice and Bob with $\sigma_A$ and $\sigma_B$ optimized (black lines). For comparison, we include the original Gaussian MDI protocol with optimal parameters (red lines). The solid lines correspond to the pure-loss protocols with ideal parameters $\eta=1$ and $\beta=1$, while the dashed lines correspond to a realistic scenario in which $\epsilon=0.05$, $\eta=0.98$ and $\beta=0.95$.}
	\label{fig:TBEN}
\end{figure}

Fig.~\ref{fig:RatesBoth} shows the total-distance between Alice and Bob as a function of the rates of all variations of the protocol in the symmetric configuration ($\tau_A=\tau_B$) and assuming a pure-loss attack ($\epsilon=\epsilon_A=\epsilon_B=0$) with perfect detection efficiency. The rates are optimized over the variances $\sigma_A$ and $\sigma_B$ ($\sigma_A$ and $\mu$ for restricted eavesdropping). For comparison we include the rate of the original Gaussian MDI protocol \cite{pirandola2015high} with equivalent parameters. At the cost of a lower rate at short distances, our protocol improves the range at which the parties may communicate. It is important to note that a fully secure rate in which Bob broadcasts less information may lie anywhere between the rates of the complete and restricted eavesdropping cases, but despite being the worst-case scenario, the rate under complete eavesdropping offers a notable advantage over the Gaussian MDI protocol.

\begin{figure}[t!]
\includegraphics[width=\linewidth]{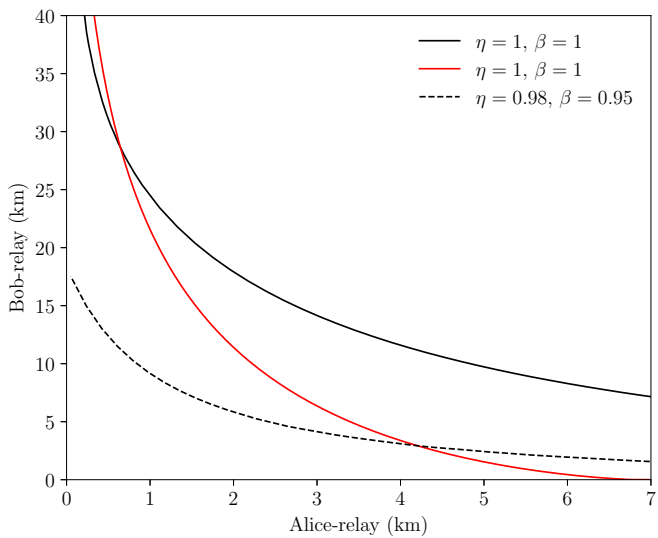}
\caption{Comparison of the maximum Bob-relay distance as a function of the Alice-relay distance under complete eavesdropping. The black lines represent our protocol with the solid line corresponding to the pure-loss case with ideal parameters $\eta=1$ and $\beta=1$ and the dashed line corresponding to case with $\epsilon=0.05$ and imperfect parameters $\eta=0.98$ and $\beta=0.95$. For comparison, the red line represents the pure-loss Gaussian MDI protocol with ideal parameters.}
\label{fig:distdist}
\end{figure}

Fig.~\ref{fig:TBEN} shows rates of protocol under complete eavesdropping as a function of the total distance between Alice and Bob. We show the pure-loss rate with ideal parameters $\eta=1$ and $\beta=1$ as well as a realistic rate with excess noise $\epsilon=0.05$, detector efficiency of 98\% and reconciliation efficiency of 95\%. Again, we also show the optimal rates of the Gaussian MDI protocol with identical parameters. Our protocol provides an advantage over the original MDI protocol under ideal as well as realistic parameters. In Fig.~\ref{fig:distdist} we explore the asymmetric configuration of the protocol under complete eavesdropping. We see that our protocol offers the biggest advantage as the symmetry of the configuration increases. However, we still observe an advantage in the asymmetric regime up to very asymmetric configurations with less than \SI{1}{\km} separating Alice from the relay.

\begin{figure}[ht]
\includegraphics[width=\linewidth]{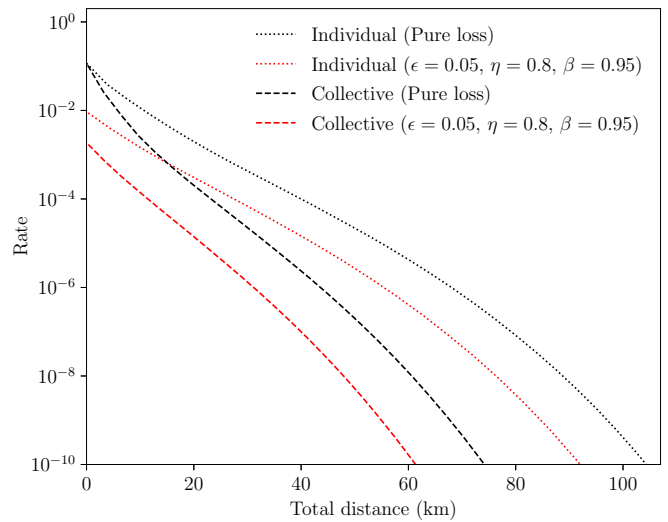}
\caption{Rates of the symmetric protocol with restricted eavesdropping as a function of the total distance between Alice and Bob with $\sigma_A$ and $\mu$ optimized. The black lines correspond to the pure-loss case with perfect detection and reconciliation while the red lines represent the rate with parameters $\epsilon=0.05$, $\eta=0.8$, and $\beta=0.95$.}
\label{fig:ratesPB}
\end{figure}

To explore the effect of the realistic parameters in more detail, we consider in Fig.~\ref{fig:ratesPB}, for individual and collective attacks with restricted eavesdropping, the rates with $\epsilon=0.05$, $\eta=0.8$ and $\beta=0.95$ (these are typical choices~\cite{Zhou19}) in the symmetric configuration. For each rate, we have incorporated $\eta$ by scaling the transmissivities on each link. This has a considerable effect on the rate but a distance exceeding \SI{60}{\km} with collective attacks is still possible. In appendix \ref{ExpOpts}, we consider the optimal parameters $\sigma_A$ and $\mu$ for an experimental configuration.

\section{Conclusions}
In this work, we have introduced a long-distance CV MDI QKD protocol with a general mathematical formulation with collective attacks which can include excess noise and experimental inefficiencies. We have demonstrated that our protocol surpasses the range of the original Gaussian CV MDI QKD protocol in both symmetric and asymmetric configurations. This improvement exists in the most powerful eavesdropping scenario and is substantially increased to distances exceeding 50 km if restricted eavesdropping is considered with either individual or collective attacks. In future work, it would be beneficial to explore an achievable fully-secure rate between these extremes if Bob can communicate all of the necessary information to Alice without broadcasting the absolute value of his measurement in each use of the protocol.

 Our protocol is robust against excess noise as well as detection and reconciliation inefficiencies and it is, therefore, a significant step towards a realistic experimental implementation. We have demonstrated that CV MDI QKD need not be restricted to short distances. In fact, our protocol provides a theoretical framework for MDI QKD at distances previously achievable only with discrete variable protocols, obtainable with inexpensive and easily implementable equipment.

\bigskip

\textbf{Acknowledgments}. This work has been funded by the European Union via ``Continuous Variable Quantum Communications'' (CiViQ, grant agreement No 820466) and the EPSRC via the `Quantum Communications hub' (EP/M013472/1, EP/T001011/1). SP would like to thank Y.-C. Zhang and X.-B. Wang for their useful suggestions about previous literature.

\appendix

\section{Optimal parameters\label{ExpOpts}}
\begin{figure}[h!]
\includegraphics[width=\linewidth]{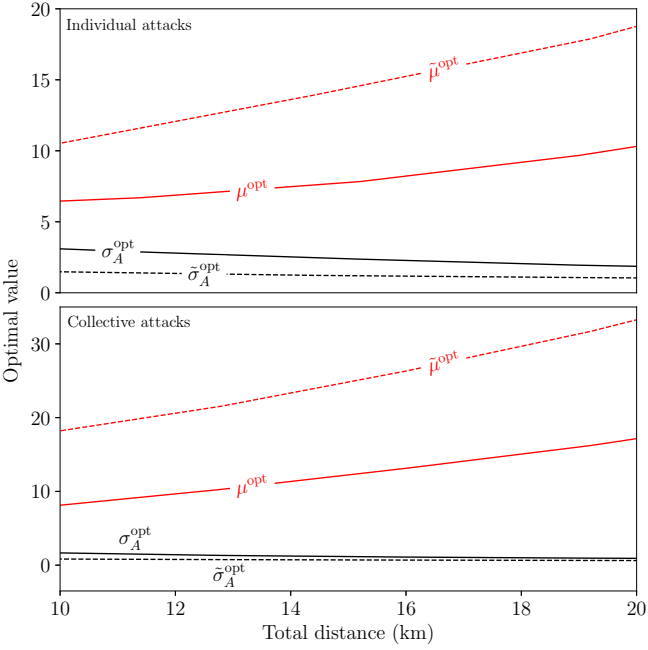}
\caption{Optimal values of $\mu$ (red lines) and $\sigma_A$ (black lines) for the symmetric protocol with restricted eavesdropping under individual (top panel) and collective (bottom panel) attacks. The solid lines represent the optimal parameters for the pure-loss case with ideal detection efficiency and the dashed lines represent the optimal values under parameters $\epsilon=0.05$, $\eta=0.8$ and $\beta=0.95$. }
\label{fig:PBOptimal}
\end{figure}

For the purposes of a proof-of-concept experiment, we show in Fig.~\ref{fig:PBOptimal} the optimal values of the parties' free parameters $\sigma_A$ and $\mu$ for the symmetric protocol with restricted eavesdropping under individual (top) and collective (bottom) attacks in a distance window of 10-\SI{20}{\km}. In both cases, we consider both pure loss with ideal reconciliation efficiency as well as a sub-optimal parameters of $\epsilon=0.05$, $\eta=0.8$ and $\beta=0.95$, matching the regimes considered in Fig.~\ref{fig:ratesPB}. For the former, we label the optimal parameters $\sigma_A^\text{opt}$ and $\mu^\text{opt}$ while for the latter we use $\tilde{\sigma}_A^\text{opt}$ and $\tilde{\mu}^\text{opt}$. We note that the optimal parameters are small relative to the original Gaussian MDI protocol in which the optimal value of the modulation tends to infinity in the case of unit reconciliation efficiency. 

\section{Entropy approximation of a non-Gaussian state \label{GEA}}
To avoid complex treatment of non-Gaussian states in the Fock basis, we will introduce an approximation for the entropy of a particular type of non-Gaussian state that is composed of the average of two Gaussian states with the same CM and different mean values. We use the CM and mean values of the constituent states to write a formula for the CM of the total state, then, by treating it as Gaussian, we use this CM to estimate its entropy. This approximation is most accurate for states with small higher-order moments, but the Gaussian assumption ensures that it is an upper bound on the entropy of any state of this form. This fact makes the approximation particularly useful in quantum key distribution when calculating the total entropy of an eavesdropper's non-Gaussian state in the Holevo bound.

We will label the constituent states of the global state $\hat{\rho}$ as $\hat{\rho}_+$ and $\hat{\rho}_-$ with associated probabilities $p(+)$ and $p(-)$, respectively. The general non-Gaussian state can then be written as
\begin{equation}
	\hat{\rho} =\sum_{\kappa =\pm} p(\kappa) \hat{\rho}_\kappa.
\end{equation}
 Let us now recall the definitions of the mean value and CM of a Gaussian state $\hat{\rho}$. The mean value of an operator $\hat{x}_i$ for a state $\hat{\rho}$ is given by
\begin{equation}
	\bar{x}_i = \langle \hat{x}_i\rangle = \tr({\hat{x}_i\hat{\rho}})
	\label{meanValueGeneral}
\end{equation}
and the covariance matrix of a state is given by
\begin{equation}
	V_{ij} = \frac{1}{2}\langle \{ \Delta \hat{x}_i, \Delta \hat{x}_j\}\rangle =\frac{1}{2} \tr\left[\{ \hat{x}_i, \hat{x}_j\}\hat{\rho}\right] -\bar{x}_i\bar{x}_j.
	\label{CMFormulaExpanded}
\end{equation}
Using Eq.~(\ref{CMFormulaExpanded}), we can express the elements $V_{ij}$ of the CM, $\mathbf{V}$ of a constituent state $\hat{\rho}_\kappa$ with mean value $\bar{\mathbf{x}}^\kappa$ as
\begin{equation}
	V_{ij}^\kappa + \bar{x}_i^\kappa\bar{x}_j^\kappa = \frac{1}{2} \tr\left[\{ \hat{x}_i, \hat{x}_j\}\hat{\rho}_\kappa\right],
	\label{condVij}
\end{equation}
 and we can also write the elements $V_{ij}'$ of the CM $\mathbf{V}'$ of the total state $\hat{\rho}$ as
\begin{align}
	V_{ij}' &= \frac{1}{2}\tr\left[\{\hat{x}_i, \hat{x}_j\} \left(\sum_{\kappa=\pm} p(\kappa) \hat{\rho}_\kappa\right) \right] - \bar{x}_i\bar{x}_j \nonumber \\
	&= \sum_{\kappa=\pm} p(\kappa)\frac{1}{2}\tr \left[\{\hat{x}_i, \hat{x}_j\}\hat{\rho}_\kappa \right] - \bar{x}_i\bar{x}_j.
\end{align}
We then substitute into this expression the right hand side of Eq.~(\ref{condVij}) to obtain
\begin{align}
	V_{ij}' &= \sum_{\kappa=\pm} p(\kappa) \left( V_{ij}^\kappa + \bar{x}_i^\kappa \bar{x}_j^\kappa\right) - \bar{x}_i \bar{x}_j \nonumber \\
	&= V_{ij} + \sum_{\kappa=\pm} p(\kappa)\bar{x}_i^\kappa \bar{x}_j^\kappa - \bar{x}_i\bar{x}_j,
	\label{CMElementsTotalState}
\end{align}
where we have made use of the requirement that the CMs of the constituent states are identical. Now by writing the mean values as $\bar{x}_i = \tr(\hat{x}_i \hat{\rho})=\sum_\kappa p(\kappa)\tr(\hat{x}_i\hat{\rho}_k)$,
and substituting into Eq.~(\ref{CMElementsTotalState}), we obtain
\begin{equation}
	V_{ij}'=V_{ij} + \sum_{\kappa=\pm } p(\kappa)\bar{x}_i^\kappa\bar{x}_j^\kappa-\sum_{\kappa=\pm } \sum_{\kappa'=\pm } p(\kappa)p(\kappa')\bar{x}_i^\kappa \bar{x}_j^{\kappa'}
\label{Vijpfirst}
\end{equation}
and by factoring out one of the sums we obtain
\begin{align}
	V_{ij}'&= V_{ij} + \sum_{\kappa=\pm } p(\kappa)\left[\bar{x}_i^\kappa\bar{x}_j^\kappa - \sum_{\kappa'=\pm } p(\kappa')\bar{x}_i^\kappa \bar{x}_j^{\kappa'}\right]\nonumber \\
	&= V_{ij} + \sum_{\kappa=\pm } p(\kappa)\left[\bar{x}_i^\kappa\bar{x}_j^\kappa - p(\kappa)\bar{x}_i^\kappa \bar{x}_j^\kappa - p(-\kappa)\bar{x}_i^\kappa \bar{x}_j^{-\kappa}\right]\nonumber \\
	&= V_{ij} + \sum_{\kappa=\pm } p(\kappa) p(-\kappa)\bar{x}_i^\kappa \left( \bar{x}_j^\kappa - \bar{x}_j^{-\kappa}\right),
\end{align}
where we have used $1-p(\kappa)=p(-\kappa)$. Now note that $p(\kappa)p(-\kappa)=p(+)p(-)$ for either value of $\kappa$, and $\sum_\kappa \bar{x}_i^\kappa(\bar{x}_j^\kappa - \bar{x}_j^{-\kappa})= (\bar{x}_j^+ - \bar{x}_j^-)\sum_\kappa \kappa\bar{x}_i^\kappa$. Therefore we obtain
\begin{align}
	V_{ij}' &= V_{ij}^+ + p(+)p(-)(\bar{x}_j^+ - \bar{x}_j^-)\sum_{\kappa=\pm} \kappa \bar{x}_i^\kappa \nonumber \\
	&= V_{ij}^+ + p(+)p(-)(\bar{x}_j^+ - \bar{x}_j^-)\left(\bar{x}_i^+ - \bar{x}_i^-\right).
	\label{SingleSumGEA}
\end{align}
We can write this in compact outer product form as
\begin{equation}
	\mathbf{V}' = \mathbf{V} + p(+)p(-)\Delta\bar{\mathbf{x}}\cdot\Delta\bar{\mathbf{x}}\tran,
\end{equation}
where $\Delta\bar{\mathbf{x}} = \bar{\mathbf{x}}^+ - \bar{\mathbf{x}}^-$.

\end{document}